\documentclass[11pt]{article}
\usepackage{amssymb,amsmath,epsfig}
\def\one{{{{\rm 1} \kern -.19em {\rm l}}}}
\def\C{{{{\rm {\mbox{\small l}}} \kern -.50em {\rm C}}}}
\def\R{{{{\rm l} \kern -.15em {\rm R}}}}
\def\N{{{{\rm l} \kern -.15em {\rm N}}}}
\def\E{{{{\rm l} \kern -.15em {\rm E}}}}
\def\P{{{{\rm l} \kern -.15em {\rm P}}}}
\def\Z{{{{\rm Z} \kern -.35em {\rm Z}}}}
\def\1{{{{\rm 1} \kern -.35em {\rm 1}}}}

\begin{document}
\begin{sloppypar}
\vspace*{0cm}
\begin{center}
{\setlength{\baselineskip}{1.0cm}{ {\Large{\bf
Construction of zero energy states in graphene through the\\ supersymmetry formalism\\}} }}
\vspace*{1.0cm}
{\large{\sc{Axel Schulze-Halberg}}}$^\dagger$ and {\large{\sc{Pinaki Roy}}}$^\ddagger$
\end{center}
\noindent \\

$\dagger$ Department of Mathematics and Actuarial Science and Department of Physics, Indiana University Northwest, 3400 Broadway,
Gary IN 46408, USA, e-mail: axgeschu@iun.edu, xbataxel@gmail.com \\ \\

$\ddagger$ Physics and Applied Mathematics Unit, Indian Statistical Institute, Kolkata 700108, India, 
e-mail: pinaki@isical.ac.in \\ \\

\vspace*{.5cm}
\begin{abstract}
\noindent
We devise a supersymmetry-based method for the construction of zero-energy states in graphene. 
Our method is applied to a two-dimensional massless Dirac equation with a hyperbolic scalar potential. We 
determine supersymmetric partners of our initial system and derive a reality condition for the transformed 
potential. The Dirac potentials generated by our method can be used to approximate interactions that are 
experimentally realizable.

\end{abstract}


\section{Introduction}
Graphene is an atomically thin conducting material that consists of carbon atoms forming a honeycomb lattice 
structure. Ever since its isolation \cite{novo}, graphene has been subject to intensive research, leading to the 
discovery of many unusual properties. One of these properties is graphene's very high electric conductivity, where 
both electrons and holes serve as charge carriers \cite{neto}. The low-energy electronic states in graphene 
can be described by the two-dimensional Dirac equation for massless particles \cite{gonz} (emergent Dirac fermions). In order to 
control the motion of the charge carriers, electromagnetic fields 
\cite{demart} \cite{dellanna} \cite{khei} \cite{toke} or scalar potentials \cite{stone} \cite{hartmann} \cite{cheianov} \cite{pereira} 
\cite{downing} can be employed. While for the vast majority of 
such fields or potentials the Dirac equation will not render solvable, there are some exceptions. 
Particular work on such exceptional cases includes studies on quasi-bound state solutions of the Dirac equation 
in a magnetic quantum dot \cite{masir}, spectrally isomorphic Dirac systems modeling graphene in an electromagnetic 
field \cite{jak}, zero-energy states in graphene under the presence of 
magnetic fields \cite{check} \cite{roy} and scalar potentials \cite{ghosh} \cite{ho}, among 
others. In the present note, we will devise a method to generate scalar potentials for which the two-dimensional 
Dirac equation admits zero-energy states. We relate our problem to a one-dimensional scenario by imposing the 
condition that our Dirac potential depends on a single coordinate only. This allows for the use of methods that 
are applicable to one-dimensional quantum systems. Such an approach has been successfully taken in previous works. 
Recent examples include modeling of Dirac fermion confinement within graphene using specific potentials \cite{ghosh} 
\cite{jak2}, a study of Dirac equations that feature periodic potentials and ${\cal{PT}}$-symmetry \cite{correa}, among 
many others. In the present work we start out from a specific hyperbolic scalar potential suitable for 
electron confinement that was 
studied in \cite{ghosh}, we will apply the quantum-mechanical supersymmetry (SUSY) formalism in order to generate 
solvable cases of the two-dimensional Dirac equation at zero energy. This formalism is one of the most 
effective methods for the generation of solvable quantum models. Based
on the mathematical concept of Darboux transformations that were first introduced in \cite{darboux},
the SUSY formalism interrelates quantum systems (SUSY partners) by means of linear differential operators
(SUSY transformation). Since there
is a vast amount of literature on the topic that encompasses many applications to particular
quantum models, we refer the reader to the self-contained reviews \cite{cooper} \cite{djf} \cite{junker} and references
therein. The remainder of this work is organized as follows. In section 2 we construct a general 
SUSY transformation between two-dimensional Dirac equations, while section 3 is devoted to the reality condition for the 
transformed Dirac potential. In section 4 we introduce the initial Dirac system 
that our SUSY-based construction will be applied to. Section 5 contains the actual calculations of the SUSY partners to 
our initial system, including the derivation of a reality condition for the SUSY-transformed potentials.

\section{The Dirac-SUSY formalism}
We start out from the two-dimensional, massless stationary 
Dirac equation, taken at zero energy. Upon employing atomic units, this equation is of the form
\begin{eqnarray}
\left[\sigma_1~p_x+\sigma_2~p_y + V(x) \right] \Psi(x,y) &=& 0, \label{dirac}
\end{eqnarray}
where the Fermi velocity was set equal to one, $\sigma_j$, $j=1,2$, denote the Pauli spin matrices and $p_x, p_y$ stand for the respective momentum 
operators. Furthermore, $V$ is the potential, and 
$\Psi$ denotes the solution spinor. We will now devise a scheme for the construction of solutions to our Dirac 
equation (\ref{dirac}) by means of the SUSY formalism. Our description of the procedure is visualized in diagram 1. 
In particular, the Dirac equation (\ref{dirac}) is abbreviated as DE in the top left corner in the diagram.
\vspace{-.5cm}
\begin{center}
\setlength{\unitlength}{1.5cm} 
\begin{picture}(4,5)
\put(-0.3,4.05){DE}
\put(3.8,4.05){DE$'$}
\put(1.1,1.25){{\scriptsize{SUSY transformation}}}
\put(.75,4.25){{\scriptsize{Dirac SUSY transformation}}}
\put(.5,4.15){\vector(1,0){3}}
\put(-0.1,1.05){SE}
\put(.15,3.75){\vector(0,-1){2.25}}
\put(-.45,2.7){\scriptsize{Point}}
\put(-1.25,2.5){\scriptsize{transformation}}
\put(3.8,1.05){SE$'$}
\put(3.85,1.5){\vector(0,1){2.25}}
\put(4.05,2.7){\scriptsize{Point}}
\put(4.05,2.5){\scriptsize{transformation}}
\put(.5,1.15){\vector(1,0){3}}
\end{picture} \\[0pt]
\vspace{-1cm} Diagram 1: Dirac equation and SUSY formalism
\end{center}
\vspace{0.5cm} 
Since the 
potential depends only on the variable $x$, but not on $y$, we can represent $\Psi$ in the following way:
\begin{eqnarray}
\Psi(x,y) &=& \frac{1}{2}~\exp\left(i~k_y~y \right)
\left(
\begin{array}{ll}
\Psi_1(x)+\Psi_2(x) \\[1ex]
\Psi_1(x)-\Psi_2(x)
\end{array}
\right), \label{psinew}
\end{eqnarray}
where the wave number $k_y$ describes free motion in the $y$-direction. Upon substitution of the point transformation
(\ref{psinew}) into (\ref{dirac}), we obtain a system of equations for the functions $\Psi_1$ and 
$\Psi_2$. This system has the form
\begin{eqnarray}
-i~\Psi_1'(x)+V(x)~\Psi_1(x) +i~k_y~\Psi_2(x) &=& 0 \label{syst1} \\[1ex]
i~\Psi_2'(x)+V(x)~\Psi_2(x) -i~k_y~\Psi_1(x) &=& 0. \label{syst2}
\end{eqnarray}
After decoupling this system \cite{ghosh}, we arrive at the following relations
\begin{eqnarray}
\Psi_1''(x)+\Big[V(x)^2+i~V'(x)-k_y^2\Big] \Psi_1(x) ~=~0 \label{eq1} 
\end{eqnarray}
\vspace{-.5cm}
\begin{eqnarray}
\hspace{-.85cm} \Psi_2(x) ~=~ \frac{1}{k_y} ~\Big\{\Psi_1'(x)+i~V(x) \Psi_1(x)\Big\},  \label{psi2}
\end{eqnarray}
where we assume that $k_y \neq 0$. Note that this assumption is not restrictive: if $k_y$ vanishes, the system 
(\ref{syst1}), (\ref{syst2}) is not coupled anymore, such that $\Psi_1$ and $\Psi_2$ can be determined directly:
\begin{eqnarray}
\Psi_1(x) ~=~ \exp\left[-i \int\limits^x V(t)~dt \right] \qquad \qquad
\Psi_2(x) ~=~ \exp\left[i \int\limits^x V(t)~dt \right]. \nonumber
\end{eqnarray}
Upon substituting these functions into the spinor (\ref{psinew}) we obtain the norm $\Vert \Psi \Vert =~$constant, which 
is not interesting here. We can therefore assume that $k_y \neq 0$. Now, equation (\ref{eq1}) has the form of a Schr\"odinger equation (SE, lower left corner of our diagram). 
Consequently, we can solve our Dirac equation (\ref{dirac}) by determining its first solution component 
from (\ref{eq1}) and afterwards evaluating (\ref{psi2}) in order to get the second solution component. 
Now, we are interested in generating potentials $V$, for which the Dirac equation (\ref{dirac}) has closed-form 
solutions. This can be achieved by applying the SUSY formalism to (\ref{eq1}), since it has the form of 
a Schr\"odinger equation. 
To this end, let us assume that  $\Psi_1$ solves our Schr\"odinger-type equation (\ref{eq1}).
We introduce functions $u_1,u_2,...,u_n$ that are 
solutions to (\ref{eq1}), such that the set $\{u_1,...,u_n,\Psi_1\}$ is linearly independent. 
Typically, linear independence is achieved by associating the transformation functions with different values of $-k_y^2$, 
usually referred to as factorization energies that in this work will be labeled $\epsilon_1,...,\epsilon_n$.
The $n$-th order SUSY transformation of $\Psi_1$ has the form
\begin{eqnarray}
\Phi_1(x) &=& \frac{W_{u_1,...,u_n,\Psi_1}(x)}{W_{u_1,...,u_n}(x)}, \label{susy}
\end{eqnarray}
where $W$ stands for the Wronskian of the functions in its index. 
It can be shown \cite{bagrov} \cite{xbatcon} that our function (\ref{susy}) 
satisfies the equation
\begin{eqnarray}
\Phi_1''(x)+\Big[\hat{V}(x)-k_y^2\Big]~ \Phi_1(x) ~=~0, \label{eqt1} 
\end{eqnarray}
for a transformed potential $\hat{V}$ that has the form
\begin{eqnarray}
\hat{V}(x) &=& V(x)^2+i~V'(x)+2~\frac{d^2}{dx^2} ~\log\left[W_{u_1,...,u_n}(x) \right]. \label{pott}
\end{eqnarray}
The remaining task is to find the potential for the transformed Dirac equation. Even though we are given the explicit form 
(\ref{pott}) of the Schr\"odinger-type potential in (\ref{eqt1}), we must match it with its counterpart in (\ref{eq1}) 
in order to obtain the Dirac potential. This implies that we must find a function $U$ satisfying the condition
\begin{eqnarray}
U(x)^2+i~U'(x) &=&V(x)^2+i~V'(x) +2~\frac{d^2}{dx^2} \log\left[W_{u_1,...,u_n}(x) \right]. \label{potcon}
\end{eqnarray}
The general solution to this Riccati equation can be constructed according to a known procedure \cite{kamke}, based on 
the linearization of (\ref{potcon}) by substituting
\begin{eqnarray}
U(x) &=& i~\frac{\hat{\Phi}_1'(x)}{\hat{\Phi}_1(x)}. \nonumber
\end{eqnarray}
This setting converts (\ref{potcon}) into
\begin{eqnarray}
\hat{\Phi}_1''(x)+\Big\{V(x)^2+i~V'(x) +2~\frac{d^2}{dx^2} \log\left[W_{u_1,...,u_n}(x) \right] \hspace{-.1cm}\Big\}~ 
\hat{\Phi}_1(x) ~=~0. \label{zero}
\end{eqnarray}
We observe that this equation matches (\ref{eqt1}) for $k_y=0$. Consequently, we obtain a solution of 
(\ref{zero}) by applying the setting $k_y=0$ to a solution (\ref{susy}) of (\ref{eq1}), that is, we set
\begin{eqnarray}
\hat{\Phi}_1(x) &=& \Phi_1(x)_{|k_y=0}. \label{phi1zero}
\end{eqnarray}
Since the solution $\Phi_1$ of (\ref{eqt1}) will depend on $k_y$, we simply set it to zero in order to get $\hat{\Phi}_1$. 
Now, using the latter function, we can state the general solution to (\ref{potcon}) in the form
\begin{eqnarray}
U(x) &=& i~\frac{\hat{\Phi}_1'(x)}{\hat{\Phi}_1(x)}+\frac{i~K}{C~\hat{\Phi}_1(x)^2+\hat{\Phi}_1(x)^2~
\displaystyle{\int\limits^x \frac{1}{\hat{\Phi}_1(t)^2}~dt}}, \label{u}
\end{eqnarray}
where $K \in \{0,1\}$ and $C$ is an arbitrary constant. Using (\ref{potcon}) and (\ref{u}), 
our equation (\ref{eqt1}) now reads
\begin{eqnarray}
\Phi_1''(x)+\Big[U(x)^2+i~U'(x)-k_y^2\Big]~ \Phi_1(x) ~=~0, \label{set}
\end{eqnarray}
note that for the sake of brevity we did not include the full form (\ref{u}) of $U$. Similar to its counterpart (\ref{eq1}), 
equation (\ref{set}) is of Schr\"odinger type (SE$'$, lower right corner of the diagram). According to (\ref{psi2}), we can 
construct the counterpart $\Phi_2$ of $\Phi_1$ through the identity
\begin{eqnarray}
\Phi_2(x) &=& \frac{1}{k_y} ~\Big\{\Phi_1'(x)+i~U(x) \Phi_1(x)\Big\}.  \label{phi2}
\end{eqnarray}
In the final step we plug (\ref{susy}) and (\ref{phi2}) into the partner point transformation of (\ref{psinew}), that is, 
\begin{eqnarray}
\Phi(x,y) &=& \frac{1}{2}~\exp\left(i~k_y~y \right)
\left(
\begin{array}{ll}
\Phi_1(x)+\Phi_2(x) \\[1ex]
\Phi_1(x)-\Phi_2(x)
\end{array}
\right). \nonumber
\end{eqnarray}
Now, the function $\Phi$ provides a solution of the zero-energy Dirac equation (DE$'$, upper right corner of the diagram) 
\begin{eqnarray}
\left[\sigma_1~p_x+\sigma_2~p_y + U(x) \right] \Phi(x,y) &=& 0, \label{diract}
\end{eqnarray}
recall that the potential $U$ is defined in (\ref{u}). In summary, we have devised a method for constructing 
solutions of the Dirac equation (\ref{dirac}). We will refer to the transformed potential $U$ 
as the SUSY partner of $V$ in (\ref{dirac}). Observe that in its general form the latter potential is complex-valued.

\section{Reality condition for the transformed Dirac potential}
In its general form (\ref{u}), our transformed potential has a nonvanishing imaginary part. Since we are interested in 
real-valued potentials only, our SUSY transformation must be chosen such that the imaginary part in (\ref{u}) vanishes. 
To this end, we consider the first term on the right side of (\ref{u}). This term is real-valued if the following condition 
holds
\begin{eqnarray}
\mbox{Im} \left[i~\frac{\hat{\Phi}_1'(x)}{\hat{\Phi}_1(x)} \right] &=& 0. \label{realcon}
\end{eqnarray}
In order to satisfy this condition, the function $\hat{\Phi}_1$ must have nonvanishing imaginary part. In addition, 
the complex absolute value of $\hat{\Phi}_1$ is required to be constant, that is, 
\begin{eqnarray}
\left|\hat{\Phi}_1(x) \right| &=& r_1, \label{real}
\end{eqnarray}
where $r_1$ is a nonnegative number. Note that this implies existence of a real-valued function $F$, such that 
\begin{eqnarray}
\hat{\Phi}_1(x) &=& r_1~\exp\left[i~F(x) \right]. \label{phipol}
\end{eqnarray}
Now, upon substituting the definitions (\ref{susy}) and (\ref{phi1zero}) into (\ref{real}), 
our condition (\ref{real}) takes the form
\begin{eqnarray}
\frac{\left|W_{u_1,...,u_n,\Psi_1}(x)_{|k_y=0} \right|}{\left|W_{u_1,...,u_n}(x) \right|} &=& r_1. \label{realfin}
\end{eqnarray}
This constraint on our initial solution and the transformation functions states that the quotient on the left side of 
(\ref{realfin}) must be constant, in particular, it cannot depend on $x$. Note that (\ref{realfin}) is not trivial because both 
Wronskians must have nonvanishing imaginary part. Let us now assume that (\ref{realcon}) is fulfilled, 
such that the first term on the right side of (\ref{u}) is real-valued. The second term can then be rewritten by 
decomposing $\hat{\Phi}_1$ into its real and imaginary part. To this end, we first find that
\begin{eqnarray}
\frac{1}{\hat{\Phi}_1(x)^2} &=& \frac{\hat{\Phi}^\ast_1(x)^2}{\hat{\Phi}_1(x)^2~\hat{\Phi}^\ast_1(x)^2} 
~=~ \frac{\hat{\Phi}^\ast_1(x)^2}{|\hat{\Phi}_1(x)|^4} ~=~ \frac{1}{r_1^4}~\hat{\Phi}^\ast_1(x)^2, \nonumber
\end{eqnarray}
where in the last step we used our assumption (\ref{real}). We now incorporate this result in the second term on the 
right side of (\ref{u}), then multiply numerator and denominator by $\hat{\Phi}^\ast_1(x)^2$. This gives
\begin{eqnarray}
\frac{i~K}{C~\hat{\Phi}_1(x)^2+\hat{\Phi}_1(x)^2~
\displaystyle{\int\limits^x \frac{1}{\hat{\Phi}_1(t)^2}~dt}} &=& \frac{i~K}{C~\hat{\Phi}_1(x)^2+\hat{\Phi}_1(x)^2~
\displaystyle{\int\limits^x \frac{1}{r_1^4}~\hat{\Phi}^\ast_1(t)^2~dt}}\nonumber \\[1ex]
&=& \frac{i~K~\hat{\Phi}^\ast_1(x)^2}{C~|\hat{\Phi}_1(x)|^4+|\hat{\Phi}_1(x)|^4~
\displaystyle{\int\limits^x \frac{1}{r_1^4}~\hat{\Phi}^\ast_1(t)^2~dt}}\nonumber \\[1ex]
&=& \frac{i~K~\hat{\Phi}^\ast_1(x)^2}{C~r_1^4+\displaystyle{\int\limits^x \hat{\Phi}^\ast_1(t)^2~dt}}. \nonumber
\end{eqnarray}
Since the quantity $C$ is arbitrary, we can without restriction absorb it into the integral as an integration constant. 
This yields the identity
\begin{eqnarray}
\frac{i~K}{C~\hat{\Phi}_1(x)^2+\hat{\Phi}_1(x)^2~
\displaystyle{\int\limits^x \frac{1}{\hat{\Phi}_1(t)^2}~dt}} &=& i~K~\frac{\hat{\Phi}^\ast_1(x)^2}
{\displaystyle{\int\limits^x \hat{\Phi}^\ast_1(t)^2~dt}}. \nonumber
\end{eqnarray}
We require the latter expression to be real-valued, obtaining a condition similar to (\ref{realcon})
\begin{eqnarray}
\mbox{Im}\left[
i~\frac{\hat{\Phi}^\ast_1(x)^2}
{\displaystyle{\int\limits^x \hat{\Phi}^\ast_1(t)^2~dt}} \right] &=& 0. \nonumber
\end{eqnarray}
This condition is fulfilled if the integral in its denominator has constant complex absolute value, that is, if
\begin{eqnarray}
\left| \int\limits^x \hat{\Phi}^\ast_1(t)^2~dt \right| &=& r_2, \nonumber
\end{eqnarray}
for a nonnegative real number $r_2$. Consequently, we can write it in polar form that corresponds to (\ref{phipol})
\begin{eqnarray}
\int\limits^x \hat{\Phi}^\ast_1(t)^2~dt &=& r_2~\exp\left[ i~G(x) \right], \label{phic}
\end{eqnarray}
for a real-valued function $G$. The identity (\ref{phic}) is solved for the function $\hat{\Phi}^\ast_1$ by taking the 
derivative on both sides. We obtain
\begin{eqnarray}
\hat{\Phi}^\ast_1(x) &=& \sqrt{i~r_2~G'(x)}~\exp\left[ i~\frac{G(x)}{2} \right]. \label{phicpol}
\end{eqnarray}
Next, we observe that the functions in (\ref{phipol}) and (\ref{phicpol}) are complex conjugates of each other. As such, 
they must have the same absolute value. We know from (\ref{real}) that 
\begin{eqnarray}
\left|\hat{\Phi}_1(x) \right|^2 &=& r_1^2. \label{abs1}
\end{eqnarray}
Now let us calculate the absolute value by means of (\ref{phicpol}). We find
\begin{eqnarray}
\left|\hat{\Phi}^\ast_1(x) \right|^2 &=& \hat{\Phi}^\ast_1(x)~\hat{\Phi}_1(x) ~=~ r_2~G'(x). \label{abs2}
\end{eqnarray}
Since (\ref{abs1}) and (\ref{abs2}) must be equal, comparison yields after integration
\begin{eqnarray}
G(x) &=& \frac{r_1^2}{r_2}~x + g, \nonumber
\end{eqnarray}
where $g$ is a real-valued constant. We substitute this result into (\ref{phicpol}) and arrive at the explicit form
\begin{eqnarray}
\hat{\Phi}^\ast_1(x) &=& \sqrt{i}~r_1~\exp\left[ i~\left(\frac{r_1^2}{2~r_2}~x + \frac{g}{2} \right) \right]. \nonumber
\end{eqnarray}
In the final step we can now find the function $\hat{\Phi}_1$ by means of complex conjugation
\begin{eqnarray}
\hat{\Phi}_1(x) &=& \sqrt{-i}~r_1~\exp\left[ -i~\left(\frac{r_1^2}{2~r_2}~x + \frac{g}{2} \right) \right]. \label{phitest}
\end{eqnarray}
Note that the root in the latter expression can be absorbed into the exponential function, such that there is no 
contradiction with the form (\ref{phipol}). Let us now calculate the transformed Dirac potential that results from 
using (\ref{phitest}). Substitution into (\ref{u}) gives the constant potential
\begin{eqnarray}
U(x) &=& \frac{r_1^2}{2~r_2}-\frac{K~r_1^2}{r_2}, \nonumber
\end{eqnarray}
where the integral in (\ref{u}) contributes a constant that must be chosen as $-C$. Note that this is not a 
restriction due to our earlier interpretation of $C$ as a constant of integration. Now, since we are not 
interested in generating constant potentials, our only option is to choose $K=0$ in (\ref{u}). 
Our transformed Dirac potential then takes the general form that can be obtained by combining 
(\ref{susy}) and (\ref{u}) for $K=0$, that is, we have
\begin{eqnarray}
U(x) &=& i~\frac{\hat{\Phi}_1'(x)}{\hat{\Phi}_1(x)} ~=~ i\left\{\frac{W_{u_1,...,u_n}(x)}{W_{u_1,...,u_n,\Psi_1}(x)} ~\frac{d}{dx} 
\left[\frac{W_{u_1,...,u_n,\Psi_1}(x)}{W_{u_1,...,u_n}(x)}\right] \right\}_{\Big| k_y=0}, 
\label{uuse}
\end{eqnarray}
where reality of the latter expression is established by means of the condition (\ref{realfin}).

\section{The initial Dirac system}
\begin{figure}[h]
\begin{center}
\epsfig{file=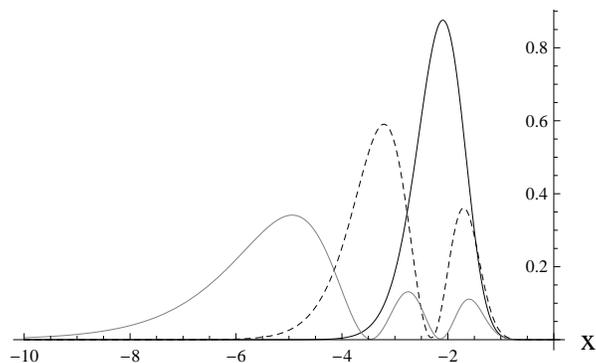,width=7.8cm}
\caption{The normalized density $|\Psi_{1,b}|^2$ associated with the solution component (\ref{psi1bound}) for the 
values $n=0$ (black curve), $n=1$ (dashed curve), and $n=2$ (gray curve), respectively.
Parameter settings are $\lambda=3$ and $\mu=10$.}
\label{inisol}
\end{center}
\end{figure}

In order to apply our method, we need 
a particular potential $V$, for which the Dirac equation (\ref{dirac}) admits closed-form solutions. 
Such a potential is given by \cite{ghosh}
\begin{eqnarray}
V(x) &=& -\lambda~\mbox{sech}(x)+\mu~\tanh(x), \label{v}
\end{eqnarray}
where $\lambda,\mu$ are real-valued constants. From a physical viewpoint, potential (\ref{v}) represents a 
well for the electrons if $\lambda>0$ and a well for the holes if $\lambda<0$ \cite{ghosh}. Solutions of our 
Dirac equation (\ref{dirac}) for the potential (\ref{v}) that are of bound-state type, therefore represent 
a confinement of these electrons or holes. We will 
distinguish these two cases below when discussing Dirac solutions of bound-state type. 
The general solution of the Dirac equation (\ref{dirac}) for the 
potential (\ref{v}) is expressed 
through relation (\ref{psinew}), where the function $\Psi_1$ is given by
\begin{eqnarray}
\Psi_1(x) &=& c_1~\cosh(x)^{-\lambda-i \mu} \left[1+i~\sinh(x) \right]^\lambda~{}_2F_1\left[
a,b,c,\frac{1}{2}-\frac{i}{2}~\sinh(x)\right]+ \nonumber \\[1ex]
& & \hspace{-1cm}+~c_2~\cosh(x)^{-\lambda+i \mu+1}~
\left[1-i~\sinh(x) \right]^{\lambda} ~{}_2F_1\left[
1-a,1-b,2-c,\frac{1}{2}-\frac{i}{2}~\sinh(x)\right]. \label{psi1x}
\end{eqnarray}
Here, $c_1, c_2$ are arbitrary constants and ${}_2F_1$ stands for the hypergeometric function \cite{abram}. 
Furthermore, the following abbreviations are in use
\begin{eqnarray}
a~=~-i~\mu+\sqrt{k_y^2-\mu^2} \qquad \qquad b~=~-i~\mu-\sqrt{k_y^2-\mu^2} \qquad \qquad c~=~\frac{1}{2}-\lambda-i~\mu. 
\nonumber
\end{eqnarray}
The function $\Psi_2$ in (\ref{psinew}) can now be obtained from (\ref{psi1x}) through the relation (\ref{psi2}). Since 
both $\Psi_2$ as well as the general solution (\ref{psinew}) take very long and involved forms, we omit to state them here. 
Let us now endow our Dirac equation (\ref{dirac}) with Dirichlet boundary conditions at the infinities, that is, 
we impose 
\begin{eqnarray}
\lim\limits_{x \rightarrow -\infty} \Psi(x) ~=~\lim\limits_{x \rightarrow \infty} \Psi(x) ~=~0. \label{bc}
\end{eqnarray}
Since both functions $\Psi_1$ and $\Psi_2$ are in general unbounded, the solution (\ref{psinew}) does not satisfy 
(\ref{bc}), unless we apply particular settings to our parameters. Before we do so, we must distinguish between 
solutions representing electrons and solutions that are associated with holes. The respective 
parameter settings are given by overall $c_2=0$ and 
\begin{eqnarray}
\begin{array}{lclllllllll}
\mbox{Electrons:} &\hspace{.1cm} \lambda ~>~ \frac{1}{2} & \hspace{.5cm} 
k_y ~=~ \pm \sqrt{\mu^2+\left(\lambda-n-\frac{1}{2} \right)^2},~~~n=0,1,2,...,\lambda-\frac{1}{2} \\[1ex]
\mbox{Holes:} & \hspace{.1cm} \lambda ~<~ -\frac{1}{2}  & \hspace{.5cm} 
k_y ~=~ \pm \sqrt{\mu^2+\left(\lambda+n+\frac{1}{2} \right)^2},~~~n=0,1,2,...,-\lambda-\frac{1}{2}.
\end{array} \label{hole}
\end{eqnarray}
Observe that these definitions of $k_y$ stem from the Schr\"odinger-type equation (\ref{eq1}). In order to 
support solutions of bound-state type, the stationary energy $\epsilon=-k_y^2$ associated with the latter equation 
must satisfy certain constraints. These constraints are precisely given by the definitions of $k_y$ in (\ref{hole}). 
Upon using the settings for representing electrons, the function (\ref{psi1x}) that determines the Dirac 
solution by means of (\ref{psinew}) and (\ref{psi2}), takes the following form 
\begin{eqnarray}
\Psi_{1,b}(x) &=& \cosh(x)^{-\lambda-i \mu} \left[1+i~\sinh(x) \right]^\lambda \times \nonumber \\[1ex]
&\times&{}_2F_1\left[
n-\lambda-i~\mu+\frac{1}{2},-n+\lambda-i~\mu-\frac{1}{2},-\lambda-i~\mu+\frac{1}{2},\frac{1}{2}-\frac{i}{2}~\sinh(x)\right], 
\label{psi1bound}
\end{eqnarray}
where $n=0,1,2,...,\lambda-\frac{1}{2}$ and some irrelevant constants have been omitted. Furthermore, 
the index $b$ indicates that $\Psi_{1,b}$ satisfies the boundary conditions (\ref{bc}). This function can also be used to represent holes if the replacement $n \rightarrow -n-1$ is made and 
the value of $\lambda$ is negative, see the 
definitions of $k_y$ in (\ref{hole}). In both cases, the respective function (\ref{psi1bound}) determines a solution 
(\ref{psinew}) that satisfies the boundary-value problem (\ref{dirac}), (\ref{bc}). Let us 
mention that (\ref{psi1bound}) can be expressed in terms of Jacobi polynomials because the 
series of the hypergeometric function in (\ref{psi1bound}) terminates after a finite number of terms. This 
might be puzzling at first sight, as the first argument of the latter hypergeometric function is not equal to a negative 
integer. We can resolve the issue by using the following identity \cite{abram}
\begin{eqnarray}
{}_2F_1(C-A,C-B,C,z) &=& (1-z)^{A+B-C}{}_2F_1(A,B,C,z), \label{hypid}
\end{eqnarray}
where $A,B,C$ and $z$ are admissible arguments. Now we compare the left side of (\ref{hypid}) with the 
hypergeometric function in (\ref{psi1bound}). We can match those two by identifying
\begin{eqnarray}
A ~=~ -n \qquad \qquad B ~=~ n-2~\lambda+1 \qquad \qquad C ~=~ -\lambda-i~\mu+\frac{1}{2}. \nonumber
\end{eqnarray}
As a consequence, identity (\ref{hypid}) applies to the hypergeometric function in (\ref{psi1bound}). The latter 
function is then converted to a form the first argument of which is the nonnegative integer $A=-n$, see right side 
of (\ref{hypid}). This implies termination of the hypergeometric series after $n+1$ terms. As a final remark let us add that it will prove 
convenient for our purposes to maintain the hypergeometric representation (\ref{psi1bound}) 
rather than replacing it through Jacobi polynomials. Figure \ref{inisol} shows examples of normalized 
probability densities associated with the solutions (\ref{psi1bound}) for a particular parameter setting.

\section{Supersymmetric partners of the initial system}
We will now apply the transformation scheme displayed in diagram 1 to our initial Dirac 
equation (\ref{dirac}) for the potential (\ref{v}), using the solutions (\ref{psi1x}). To this end, 
we will use the SUSY algorithm to construct partner potentials to (\ref{v}) with 
respect to the mapping shown in diagram 1. For the sake of brevity, subsequent calculations are restricted to the 
case of electrons, that is, the first line of (\ref{hole}).

\subsection{First-order SUSY transformations} 
In the simplest case, SUSY transformations are of first order, requiring a single transformation function for 
their application. Recall that this function must be taken from the class (\ref{psi1x}), which gives an infinite 
number of possible choices. We will distinguish here between regular transformation functions that 
satisfy the boundary conditions (\ref{bc}) and nonregular transformation functions that do not.

\begin{figure}[h]
\begin{center}
\epsfig{file=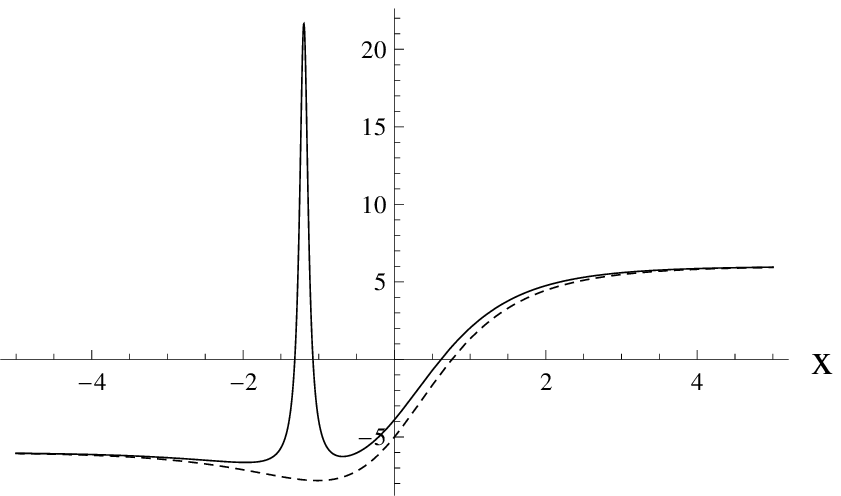,width=7.8cm}
\epsfig{file=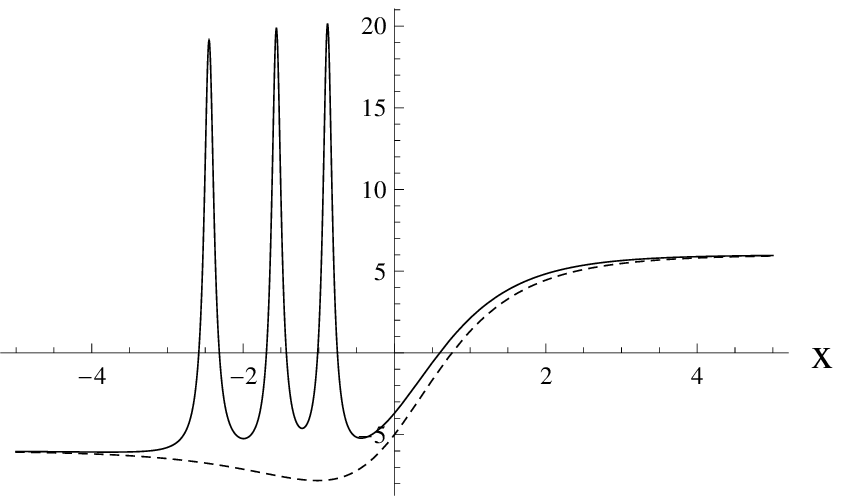,width=7.8cm}
\caption{The initial Dirac potential (\ref{v}) (dashed curve) and its 
transformed counterpart (\ref{uuse}) (solid curve), obtained from a first-order SUSY transformation using 
(\ref{psi1bound}) and the transformation functions (\ref{u1}) (left plot) and (\ref{u13}) (right plot). Overall settings are 
$\lambda=5$ and $\mu=6$. }
\label{s1pot}
\end{center}
\end{figure}

\begin{figure}[h]
\begin{center}
\epsfig{file=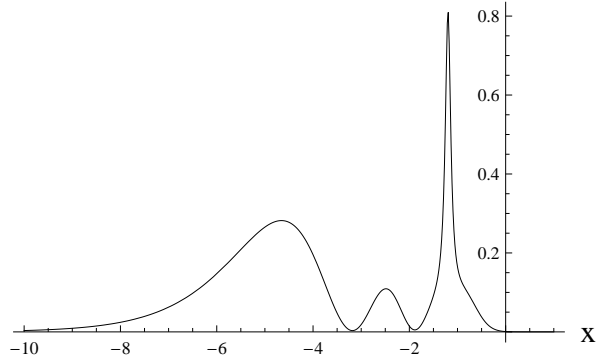,width=7.8cm}
\caption{The normalized density $|\Phi_1|^2+|\Phi_2|^2$ associated with the solution components (\ref{susy1}) and (\ref{phi2}), 
respectively, obtained from the transformation function (\ref{u1}). 
Present settings are $\lambda=5$, $\mu=6$, and $n=2$.}
\label{sol1}
\end{center}
\end{figure}

\paragraph{Regular transformation functions.} 
Transformation functions that satisfy the boundary conditions (\ref{bc}) are provided by (\ref{psi1bound}), recall that 
the parameter $n$ can take nonnegative integer values that are less than $\lambda-1/2$. Let us first choose the 
parameter value $\lambda=5$, $\mu=6$ and $n=1$ in (\ref{psi1bound}), that is,
\begin{eqnarray}
u_1(x) &=& \Psi_{1,b}(x)_{|n=1} \nonumber \\[1ex]
&=& \left[1-i~\sinh(x)\right]^{-\frac{1}{2}-3i} \left[1+i~\sinh(x)\right]^{3i}
\left[12-i+8~\sinh(x) \right] \mbox{sech}^4(x)
,  \label{u1}
\end{eqnarray}
where irrelevant overall factors were discarded. The factorization energy $\epsilon_1$ for 
the present case is obtained from the first line of (\ref{hole}). Substitution of our parameter values 
gives the explicit result
\begin{eqnarray}
\epsilon_1 &=& -\left(k_y^2 \right)_{|n=1} ~=~ -\frac{193}{4}. \nonumber
\end{eqnarray}
We will now apply our SUSY transformation (\ref{susy}) to the 
function (\ref{psi1bound}), using the transformation function (\ref{u1}). Before we do so, let us check that 
we will generate a real-valued Dirac potential by means of our transformation. To this end, we employ 
the reality condition (\ref{realfin}). Note that on the left side of our reality condition we are required to evaluate 
(\ref{psi1bound}) at $k_y=0$. Since the latter function does not depend explicitly on $k_y$, but only on $n$, 
we must determine the corresponding value of $n$, such that $k_y$ vanishes. According to (\ref{hole}), we find 
this value to be $n=\lambda-i\mu-1/2=9/2-6i$. Substitution into (\ref{realfin}) gives the result
\begin{eqnarray}
\frac{\left|W_{u_1,\Psi_{1,b}}(x)_{|k_y=0} \right|}{\left|u_1(x) \right|} ~=~
\frac{\left|W_{u_1,\Psi_{1,b}}(x)_{|n=9/2-6i} \right|}{\left|u_1(x) \right|} ~=~ 7. \nonumber
\end{eqnarray}
Since we obtain a constant, our reality condition is fulfilled, guaranteeing that the transformed potential (\ref{uuse}) 
becomes real-valued. We are now ready to apply our SUSY transformation. We plug $n=1$, (\ref{psi1x}) and (\ref{u1}) into 
(\ref{susy}) to obtain 
\begin{eqnarray}
\Phi_1(x) &=& \frac{W_{u_1,\Psi_{1,b}}(x)}{u_1(x)}, \label{susy1}
\end{eqnarray}
where we omit to state the explicit forms of (\ref{psi1bound}) and (\ref{u1}), as the resulting expressions are 
very long. Recall that the solution of our transformed Dirac equation (\ref{diract}) consists of two components, 
the first of which is computed as shown in (\ref{susy1}). The second component can be found by means of 
(\ref{phi2}). In the next step we construct the transformed Dirac potential (\ref{uuse}) by means of (\ref{phi1zero}). 
This gives
\begin{eqnarray}
U(x) &=& \left\{\frac{u_1(x)}{W_{u_1,\Psi_{1,b}}(x)} ~\frac{d}{dx} 
\left[\frac{W_{u_1,\Psi_{1,b}}(x)}{u_1(x)}\right] \right\}_{\Big| n=9/2-6i}. \label{ux1}
\end{eqnarray}
The closed form of (\ref{ux1}) is too long to be displayed here. The shape of its graph can be seen in the left 
plot of figure \ref{s1pot}. 
We observe that our 
SUSY transformation modifies the initial Dirac potential (\ref{v}) by adding a spike. Note that the spike is of finite 
height, such that our potential remains free of singularities. Modification of the parameters $\lambda$ and $\mu$ in 
the transformed potential (\ref{ux1}) changes the shape of the spike, but otherwise does not modify the 
potential qualitatively. A solution of the transformed Dirac equation (\ref{diract}) for the potential (\ref{ux1}) is shown 
in figure \ref{sol1} through its normalized probability density.
We observe from the figure that the boundary conditions (\ref{bc}) are satisfied. This behaviour 
was to be expected because both transformation function (\ref{u1}) of our SUSY transformation and the function 
(\ref{psi1bound}) it is applied to, are of bound-state type. Furthermore, we see that the 
transformed potential's spike affects the shape of the probability density at the location of the spike. 
Next, in order to understand the dependence of (\ref{ux1}) on the value of $n$, let us 
replace (\ref{u1}) by the following function
\begin{eqnarray}
u_1(x) &=& \Psi_{1,b}(x)_{|n=3} \nonumber \\[1ex]
&=& \left[1-i~\sinh(x)\right]^{-\frac{1}{2}-3i} \left[1+i~\sinh(x)\right]^{3i}
\left\{(180-15~i)~\cosh(2x)+ \right. \nonumber \\[1ex]
&+& \left.2 \left[(164-72~i)~\sinh(x)+5~\sinh(3x)
\right]
\right\}.  \label{u13}
\end{eqnarray}
The factorization energy associated with this function is obtained from (\ref{hole}). We plug in $n=3$, which gives
\begin{eqnarray}
\epsilon_1 &=& -\left(k_y^2 \right)_{|n=3} ~=~ -\frac{153}{4}. \nonumber
\end{eqnarray}
The remaining procedure of constructing the transformed Dirac potential is similar to the case $n=1$. We evaluate 
our reality condition (\ref{realfin}) for the present case. This yields
\begin{eqnarray}
\frac{\left|W_{u_1,\Psi_{1,b}}(x)_{|k_y=0} \right|}{\left|u_1(x) \right|} ~=~
\frac{\left|W_{u_1,\Psi_{1,b}}(x)_{|n=9/2-6i} \right|}{\left|u_1(x) \right|} ~\approx~ 6.18466. \nonumber
\end{eqnarray}
Since this result does not depend on $x$, the transformed Dirac potential (\ref{uuse}) will be real-valued. In particular, 
the expressions (\ref{susy1}) and (\ref{ux1}) hold for $n=3$, if the function $u_1$ is given by (\ref{u13}). The right plot in 
figure \ref{s1pot} shows a graph of the transformed potential. Inspection of the figure shows that three 
peaks of finite height were added by the SUSY transformation. This behaviour generalizes to any value of $n$ that 
is a natural number, in the sense that $n$ peaks are added to the potential. In all of these cases, the potential 
remains regular. Let us finally comment on the case $n=0$. We did not discuss the latter case because the associated SUSY 
transformation does not change the initial potential siginficantly.

\begin{figure}[h]
\begin{center}
\epsfig{file=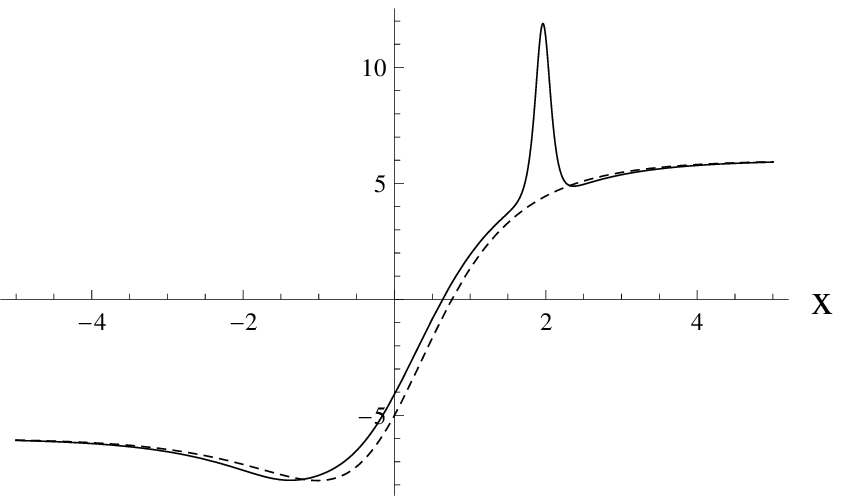,width=7.8cm}
\epsfig{file=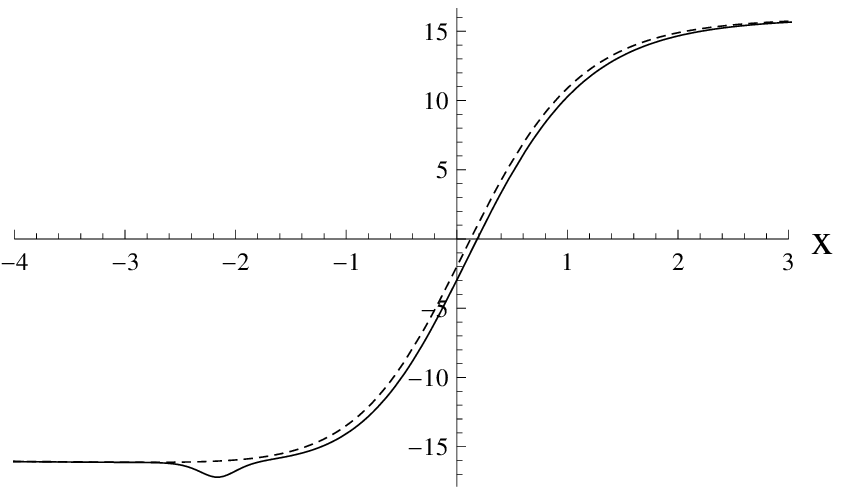,width=7.8cm}
\caption{The initial Dirac potential (\ref{v}) (dashed curve) and its 
transformed counterpart (\ref{uuse}) (solid curve), obtained from a first-order SUSY transformation using 
(\ref{psi1bound}) and the transformation functions (\ref{u10}) (left plot) and (\ref{u1m3}) (right plot). Overall settings are 
$\lambda=5$, $\mu=6$ (left plot) and $\lambda=2$, $\mu=16$ (right plot). }
\label{s1potnonreg}
\end{center}
\end{figure}

\begin{figure}[h]
\begin{center}
\epsfig{file=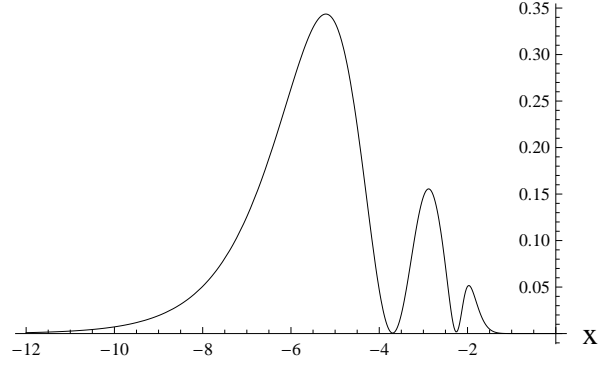,width=7.8cm}
\caption{The normalized density $|\Phi_1|^2+|\Phi_2|^2$ associated with the solution components (\ref{susy1}) and (\ref{phi2}), 
respectively, obtained from the transformation function (\ref{u1m3}). 
Present settings are $\lambda=2$, $\mu=16$, and $n=1$.}
\label{sol1n}
\end{center}
\end{figure}

\paragraph{Nonregular transformation functions.} Let us now employ first-order SUSY transformations that 
are based on transformation functions not satisfying the boundary conditions (\ref{bc}). Such functions are given by 
any particular case of (\ref{psi1x}) that does not have the form (\ref{psi1bound}). In our first example let us apply 
the parameter settings $c_1=1$, $c_2=0$, $\lambda=5$ and $\mu=6$. We obtain our transformation function in 
the following form
\begin{eqnarray}
u_1(x) &=&  \left[1-i~\sinh(x)\right]^{-\frac{1}{2}-3i} \left[1+i~\sinh(x)\right]^{3i} \mbox{sech}^4(x) \times \nonumber \\[1ex]
&\times&
{}_2F_1\left[-\frac{9}{2}-\sqrt{k_y^2-36},-\frac{9}{2}+\sqrt{k_y^2-36},-\frac{9}{2}-6~i,
\frac{1}{2}-\frac{i}{2}~\sinh(x) \right]. \label{u10}
\end{eqnarray}
When determining the remaining parameter $k_y$ in this transformation function, we must make sure that the 
transformed Dirac potential (\ref{uuse}) is real-valued. To this end, we set up our reality condition (\ref{realfin}) and 
look numerically for values of $k_y$ that render its left side constant. As a result we obtain a discrete set of values 
that is given by
\begin{eqnarray}
k_y &=& \sqrt{36+\left(\frac{9}{2}-n \right)^2},~~~n=-1,-2,-3,... \label{nreal}
\end{eqnarray}
These are precisely the values of $k_y$ defined in the first line of (\ref{hole}) when taking negative integer 
values of $n$. Any values of $k_y$ that are different from (\ref{nreal}) will result in our reality condition being 
violated. Let us therefore choose $n=-1$, which gives $k_y=\sqrt{265}/2$. The factorization energy 
$\epsilon_1=-k_y^2$ associated 
with the transformation function (\ref{u10}) is then obtained as
\begin{eqnarray}
\epsilon_1 &=& -\frac{265}{4}. \nonumber
\end{eqnarray}
Furthermore, our reality condition evaluates to
\begin{eqnarray}
\frac{\left|W_{u_1,\Psi_{1,b}}(x)_{|n=9/2-6i} \right|}{\left|u_1(x) \right|} ~\approx~ 8.13941. \nonumber
\end{eqnarray}
Since now we know that our transformed Dirac potential will be real-valued, we can perform the SUSY transformation. 
This transformation gives the potential in the form (\ref{ux1}) after substitution of (\ref{u10}) and the present 
parameter settings. The left plot in figure \ref{s1potnonreg} shows the result. We observe that 
the SUSY transformation adds a peak of finite height to the potential (\ref{v}). This behaviour does not change 
upon employing quantities $k_y$ in (\ref{nreal}) that are associated with values of $n$ different from $n=-1$. A 
scenario of several peaks that is shown in the right plot of figure \ref{s1pot}, is not feasible if we use nonregular 
transformation functions. Let us now perform another SUSY transformation, replacing (\ref{u10}) by a different 
function taken from (\ref{psi1x}). This time we use the settings $c_1=0$, $c_2=1$, $\lambda=2$ and $\mu=16$. 
The transformation function associated with these 
values takes the form
\begin{eqnarray}
u_1(x) &=&  \left[1-i~\sinh(x)\right]^{\frac{3}{2}+8i} \left[1+i~\sinh(x)\right]^{1-8i} \times \nonumber \\[1ex]
&\times&
{}_2F_1\left[\frac{5}{2}-\sqrt{k_y^2-256},\frac{5}{2}+\sqrt{k_y^2-256},\frac{7}{2}+16~i,
\frac{1}{2}-\frac{i}{2}~\sinh(x) \right]. \label{u1m3}
\end{eqnarray}
In addition, we choose the value for $k_y$ as given in (\ref{nreal}) with $n=-3$, giving $k_y=\sqrt{1105}/2$. 
The corresponding factorization energy $\epsilon_1=-k_y^2$ is then given by 
\begin{eqnarray}
\epsilon_1 &=& -\frac{1105}{4}. \nonumber
\end{eqnarray}
Upon checking our reality condition 
(\ref{realfin}) for the present case, we obtain
\begin{eqnarray}
\frac{\left|W_{u_1,\Psi_{1,b}}(x)_{|n=9/2-6i} \right|}{\left|u_1(x) \right|} ~\approx~16.6208. \nonumber
\end{eqnarray}
Since this result does not depend on $x$, our reality condition is satisfied. Upon application of our SUSY transformation 
(\ref{susy}), in combination with (\ref{ux1}), we obtain the transformed Dirac potential that is shown in 
the right plot of figure \ref{s1potnonreg}. As stated above, the SUSY transformation modifies the initial potential 
(\ref{v}) by adding a single peak. 
A typical normalized probability density of a solution to the transformed Dirac equation (\ref{diract}) 
for the potential (\ref{ux1}) is displayed in figure \ref{sol1n}. Inspection of this figure shows that the associated 
solution spinor of (\ref{diract}) complies with the boundary conditions (\ref{bc}). In particular, the probability density 
represents a bound state.

\subsection{Second-order SUSY transformations} 
We proceed by investigating the effect of second-order SUSY transformations on our Dirac potential (\ref{v}). 
Such transformations require two transformation functions that must be taken from the set (\ref{psi1x}). In the 
following we will first consider the regular case where those functions fulfill the boundary conditions (\ref{bc}).

\begin{figure}[h]
\begin{center}
\epsfig{file=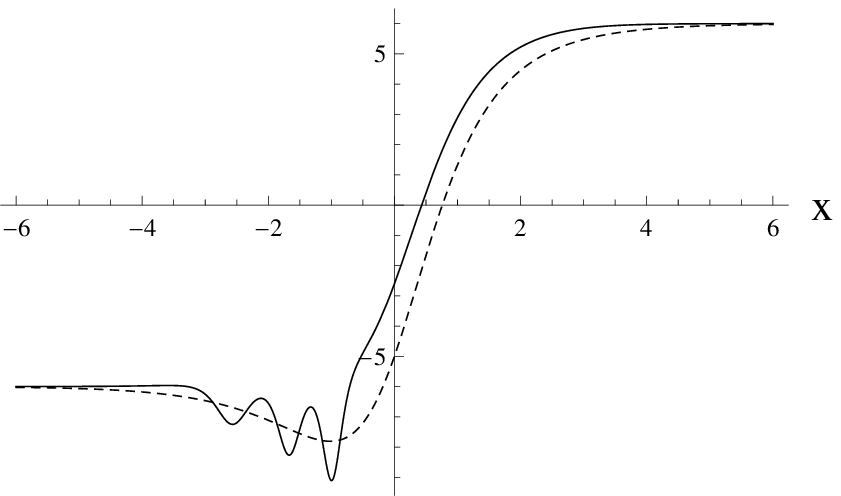,width=7.8cm}
\epsfig{file=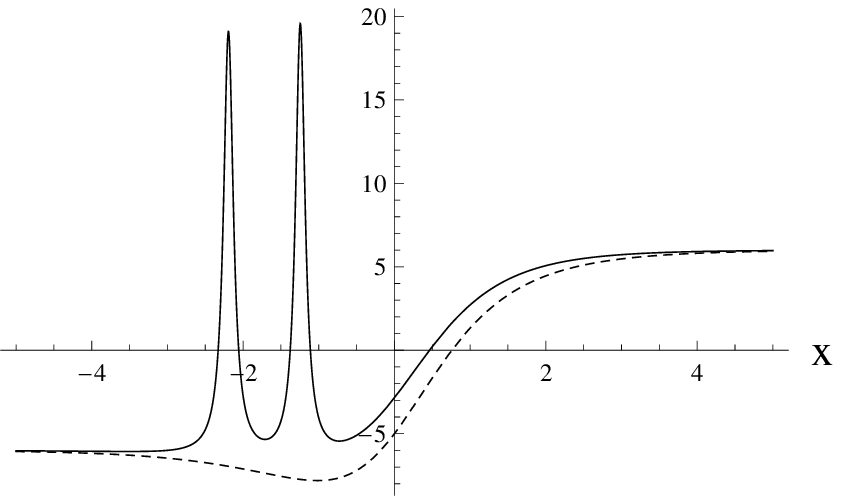,width=7.8cm}
\caption{The initial Dirac potential (\ref{v}) (dashed curve) and its 
transformed counterpart (\ref{uuse}) (solid curve), obtained from a second-order SUSY transformation using 
(\ref{psi1bound}) and the settings $\lambda=5$, $\mu=6$. Transformation functions are given by 
(\ref{u1u2b}) (left plot) and (\ref{u1u2peak2}) (right plot).}
\label{s2potr}
\end{center}
\end{figure}

\begin{figure}[h]
\begin{center}
\epsfig{file=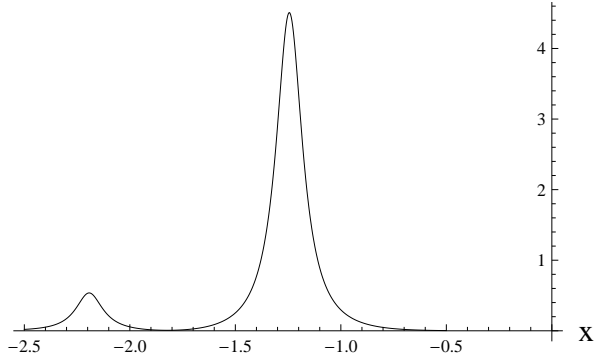,width=7.8cm}
\caption{The normalized density $|\Phi_1|^2+|\Phi_2|^2$ associated with the solution components (\ref{phi2t}) and (\ref{phi2}), 
respectively, obtained from the transformation functions (\ref{u1u2peak2}). 
Present settings are $\lambda=5$, $\mu=6$, and $n=1$.}
\label{sol2}
\end{center}
\end{figure}

\paragraph{Regular transformation functions.} In our first example we choose the same parameter values that were 
used in the first-order case. Let us apply the overall settings $\lambda=5$, $\mu=6$ to (\ref{psi1bound}). In 
addition, we pick the particular values $n=3$ and $n=4$, respectively. This gives
\begin{eqnarray}
u_1(x) &=&  \Psi_{1,b}(x)_{|n=3} \nonumber \\[1ex]
&=&  \left[1-i~\sinh(x)\right]^{-\frac{1}{2}-3i} \left[1+i~\sinh(x)\right]^{3i}
\left\{(180-15~i)~\cosh(2x)+ \right. \nonumber \\[1ex]
&+& \left.2 \left[(164-72~i)~\sinh(x)+5~\sinh(3x)
\right]
\right\} \nonumber \\[1ex]
u_2(x) &=&  \Psi_{1,b}(x)_{|n=4} \nonumber \\[1ex]
&=&  \left[1-i~\sinh(x)\right]^{-\frac{1}{2}-3i} \left[1+i~\sinh(x)\right]^{3i}
\mbox{sech}^4(x) \left\{1234+820~i+ \right. \nonumber \\[1ex]
&+& \left. (752+1024~i)~\cosh(2x)+ \right. \nonumber \\[1ex]
&+& \left. (1+2~i)~\cosh(4x)+(2088+3416~i)~\sinh(x)+(56+92~i)~\sinh(3x)
\right]. \label{u1u2b}
\end{eqnarray}
These two functions are associated with factorization energies that are obtained from (\ref{hole}). We have 
\begin{eqnarray}
\epsilon_1 &=&-\left(k_y^2\right)_{|n=3}~=~-\frac{153}{4} \nonumber \\[1ex]
\epsilon_2 &=& -\left(k_y^2\right)_{|n=4}~=~-\frac{145}{4}. \nonumber
\end{eqnarray}
Before we apply our SUSY transformation, let us verify that our reality condition (\ref{realfin}) for the 
transformed Dirac potential is satisfied. Substitution of the present parameter settings and (\ref{u1u2b}) give
\begin{eqnarray}
\frac{\left|W_{u_1,u_2,\Psi_{1,b}}(x)_{|n=9/2-6i} \right|}{\left|W_{u_1,u_2}(x) \right|} ~\approx~ 37.2366. \nonumber
\end{eqnarray}
Since this quantity is independent of the variable $x$, our transformed Dirac potential is guaranteed to take 
real values. We can now apply the second-order transformation (\ref{susy}), which reads in the present case 
of second order
\begin{eqnarray}
\Phi_1(x) &=& \frac{W_{u_1,u_2,\Psi_{1,b}}(x)}{W_{u_1,u_2}(x)}, \label{phi2t}
\end{eqnarray}
recall the the transformation functions are given by (\ref{u1u2b}). In the next step we can calculate the 
transformed potential (\ref{uuse}) by means of 
\begin{eqnarray}
U(x) &=& \left\{\frac{W_{u_1,u_2}(x)}{W_{u_1,u_2,\Psi_{1,b}}(x)} ~\frac{d}{dx} 
\left[\frac{W_{u_1,u_2,\Psi_{1,b}}(x)}{W_{u_1,u_2}(x)}\right] \right\}_{\Big| n=\frac{9}{2}-6i}. \label{ux2}
\end{eqnarray}
As before, we do not show the explicit form of the potential due to its excessive length. Instead, we refer to 
the left plot of figure \ref{s2potr} that shows the graph of (\ref{ux2}). 
We see that the SUSY transformation 
produces an oscillatory perturbation of the initial potential. This holds true for any two subsequent values of 
$n$ that enter in the regular transformation functions. In particular, variation of the parameters 
$\mu$ and $\lambda$ does not change the qualitative behaviour of the transformed potential. Let us now find out 
the effect of our second-order SUSY transformation if the two values of $n$ in the transformation functions 
$u_1$ and $u_2$ are such that there is a gap between them. To this end, we switch the value of $n$ in one of 
our transformation functions from $n=4$ to $n=0$. We have
\begin{eqnarray}
u_1(x) &=& \Psi_{1,b}(x)_{|n=0}~=~ \left[1-i~\sinh(x)\right]^{-\frac{1}{2}-3i} \left[1+i~\sinh(x)\right]^{3i} \mbox{sech}^4(x) \nonumber \\[1ex]
u_2(x) &=& \Psi_{1,b}(x)_{|n=3}, \label{u1u2peak2}
\end{eqnarray}
recall that the explicit form of the function $u_2$ can be found in (\ref{u1u2b}). The factorization energies of the 
transformation functions (\ref{u1u2peak2}) are given by (\ref{hole}), that is,
\begin{eqnarray}
\epsilon_1 &=&-\left(k_y^2\right)_{|n=0}~=~-\frac{225}{4} \nonumber \\[1ex]
\epsilon_2 &=& -\left(k_y^2\right)_{|n=3}~=~-\frac{153}{4}. \nonumber
\end{eqnarray}
We verify that our reality condition (\ref{realfin}) for the transformed Dirac potential is satisfied by substituting the 
present parameter settings and (\ref{u1u2peak2}). Evaluation gives
\begin{eqnarray}
\frac{\left|W_{u_1,u_2,\Psi_{1,b}}(x)_{|n=\frac{9}{2}-6i} \right|}{\left|W_{u_1,u_2}(x) \right|} ~\approx~ 46.3849. \nonumber
\end{eqnarray}
Since we obtain a constant, our reality condition is fulfilled. We can now proceed with the application of our second-order 
SUSY transformation (\ref{phi2t}). A normalized probability density associated with a solution of the transformed 
Dirac equation (\ref{diract}) for the present case is shown in figure \ref{sol2}.
We observe that our boundary conditions (\ref{bc}) are satisfied. Next, we determine the transformed 
Dirac potential by means of (\ref{susy}) and (\ref{ux2}). Since the result is an expression of enormous length, we omit 
to display it here, but show its graph in the right plot of figure \ref{s2potr}. Inspection of the latter plot shows that the SUSY 
transformation modifies the initial potential (\ref{v}) by inserting two peaks of finite height. This behaviour of the 
transformed potential generalizes as follows: suppose our two transformation functions $u_1$ and $u_2$ are 
associated with nonnegative integer values $n=n_1$ and $n=n_2$, respectively, such that $n_2>n_1$. As a 
consequence, the quantity $\triangle n = n_2-n_1$ is a positive integer. The second-order SUSY transformation 
will then modify the initial potential by inserting $\triangle n -1$ peaks of finite height.

\begin{figure}[h]
\begin{center}
\epsfig{file=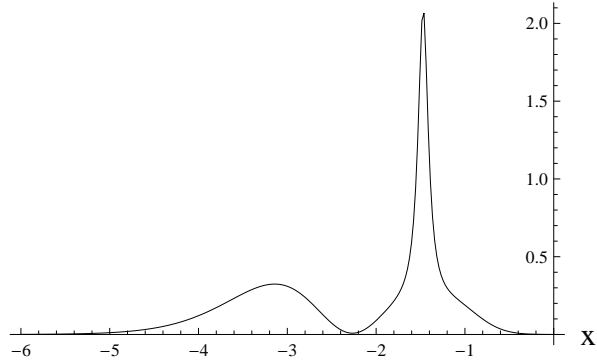,width=7.8cm}
\caption{The normalized density $|\Phi_1|^2+|\Phi_2|^2$ associated with the solution components (\ref{phi2t}) and (\ref{phi2}), 
respectively, obtained from the transformation functions (\ref{u1u2}). 
Present settings are $\lambda=5$, $\mu=6$, and $n=1$.}
\label{sol2n}
\end{center}
\end{figure}

\begin{figure}[h]
\begin{center}
\epsfig{file=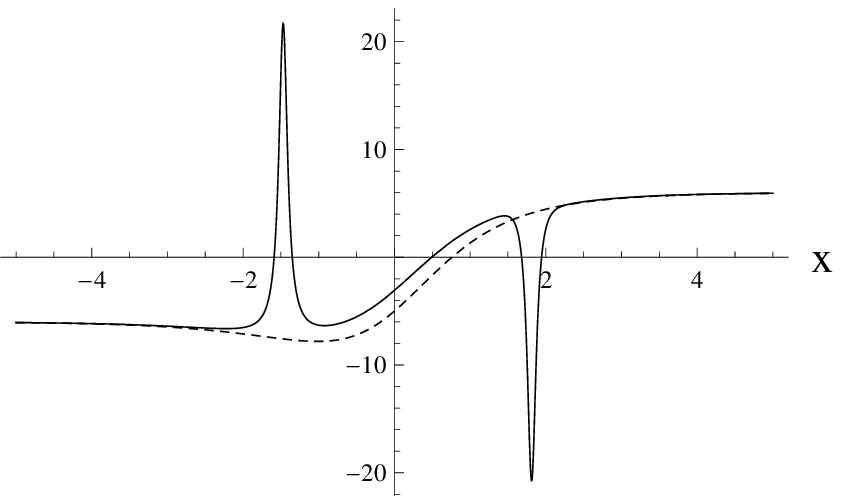,width=7.8cm}
\epsfig{file=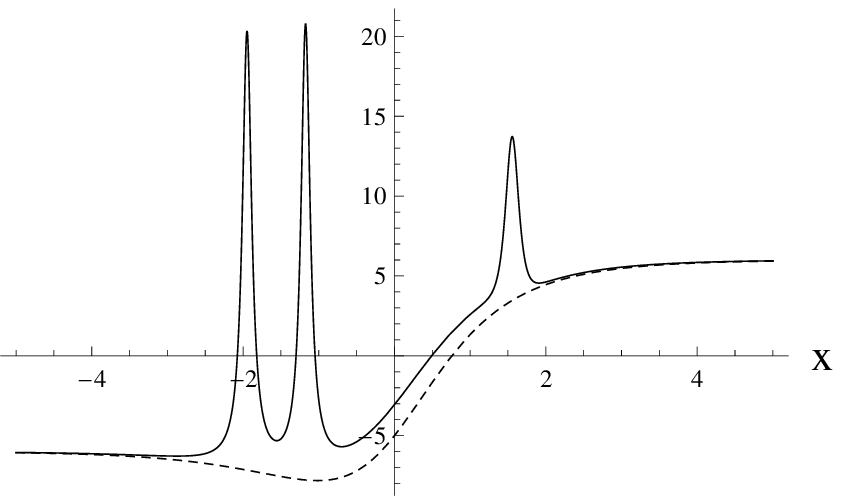,width=7.8cm}
\caption{The initial Dirac potential (\ref{v}) (dashed curve) and its 
transformed counterpart (\ref{uuse}) (solid curve), obtained from a second-order SUSY transformation using 
(\ref{psi1bound}) and the settings $\lambda=5$, $\mu=6$. Transformation functions are 
(\ref{u1u2}) (left plot) and (\ref{u1u2_13}) (right plot).}
\label{s2pot}
\end{center}
\end{figure}

\paragraph{Nonregular transformation functions.} Let us now study the behaviour of the transformed potential 
(\ref{ux2}) if one or both transformation functions are nonregular. Recall that such functions are given by the 
general solution (\ref{psi1x}), where the particular case (\ref{psi1bound}) is excluded. 
We first look at the scenario where one of the transformation functions is regular, while its counterpart is not. 
Let us apply once more the parameter settings $c_1=1$, $c_2=0$, $\lambda=5$ and $\mu=6$. In addition, we 
choose the value of $n$ associated with the regular function from (\ref{psi1bound}) as $n=1$. This gives
\begin{eqnarray}
u_1(x) ~=~ \Psi_1(x) \qquad \qquad \qquad u_2(x) ~=~ \Psi_{1,b}(x)_{|n=1}, \label{u1u2}
\end{eqnarray}
observe that the explicit form of these functions was already stated in (\ref{u10}) and (\ref{u1}), respectively. 
Before we can proceed, we must choose 
the remaining parameter $k_y$ in the function $u_1$, such that the transformed Dirac potential (\ref{uuse}) will 
be real-valued. After plugging (\ref{u1u2}) into our reality condition (\ref{realfin}), we find that the latter condition 
is fulfilled only if $k_y$ takes one of the discrete values specified in (\ref{nreal}). We choose the value 
associated with $n=-1$, such that we obtain $k_y=\sqrt{265}/2$. Now that all parameters have been assigned 
numerical values, we can specify the factorization energies for (\ref{u1u2}) from (\ref{hole}), taking into account 
our present parameter settings. We find
\begin{eqnarray}
\epsilon_1 &=&-\left(k_y^2\right)_{|n=-1}~=~-\frac{265}{4} \nonumber \\[1ex]
\epsilon_2 &=& -\left(k_y^2\right)_{|n=1}~=~-\frac{193}{4}. \nonumber
\end{eqnarray}
Our reality condition (\ref{realfin}) is satisfied, as can be seen by evaluating it for the functions (\ref{u1u2}).
\begin{eqnarray}
\frac{\left|W_{u_1,u_2,\Psi_{1,b}}(x)_{|n=\frac{9}{2}-6i} \right|}{\left|W_{u_1,u_2}(x) \right|} ~\approx~ 56.5382. \nonumber
\end{eqnarray}
As desired, we obtain a constant. Upon application of our SUSY transformation (\ref{phi2t}) we construct a solution 
to the transformed Dirac equation (\ref{diract}) for the present settings. A typical example of such a solution is 
shown in figure \ref{sol2n} by means of its probability density. 
As in the previous examples we observe from the figure that our boundary conditions (\ref{bc}) are satisfied. 
The transformed Dirac potential is now generated by application of 
(\ref{susy}) and (\ref{ux2}), resulting in a large expression. Its graph can be inspected in the left plot of 
figure \ref{s2pot}. We observe that the effect of our SUSY transformation on the initial potential (\ref{v}) is the addition of 
two finite-height peaks. In order to understand the general behaviour of the transformed potential, let us 
repeat our second-order SUSY-transformation for different transformation functions. While we maintain one of the 
functions regular and the second function nonregular, we increase the absolute values of $n$ that determine the 
factorization energies. We make the following choices in (\ref{psi1x}) and (\ref{psi1bound}), respectively
\begin{eqnarray}
u_1(x) ~=~ \Psi_1(x)_{|k_y=\frac{\sqrt{313}}{2}} \qquad \qquad \qquad u_2(x) ~=~ \Psi_{1,b}(x)_{|n=2}. \label{u1u2_13}
\end{eqnarray}
The first of these functions is nonregular. The value for $k_y$ was obtained from (\ref{nreal}) for $n=-2$, since 
this value will generate a real transformed Dirac potential. The function $u_2$ is regular. We do not state the 
explicit form of these functions because particularly $u_1$ involves long expressions. The factorization energies 
for our transformation functions can be readily calculated from (\ref{hole}). Taking into account our parameter 
values, we find
\begin{eqnarray}
\epsilon_1 &=&-\left(k_y^2\right)_{|n=-2}~=~-\frac{313}{4} \nonumber \\[1ex]
\epsilon_2 &=& -\left(k_y^2\right)_{|n=2}~=~-\frac{169}{4}. \nonumber
\end{eqnarray}
Since we intend to generate a real-valued transformed potential (\ref{uuse}), we substitute our current settings into 
the reality condition (\ref{realfin}). This gives
\begin{eqnarray}
\frac{\left|W_{u_1,u_2,\Psi_{1,b}}(x)_{|n=\frac{9}{2}-6i} \right|}{\left|W_{u_1,u_2}(x) \right|} ~\approx~ 57.4984. \nonumber
\end{eqnarray}
This is a constant, so our reality condition is satisfied. We can now construct our transformed Dirac potential 
in the form (\ref{ux2}) by substituting the present parameter values and (\ref{u1u2_13}). The graph of this 
potential is shown in the right plot of figure \ref{s2pot}. We see that the SUSY transformation modified the 
initial potential (\ref{v}) by adding three peaks of finite height. 
The general qualitative behaviour of the transformed Dirac potential can be summarized as follows. If both 
transformation functions are nonregular, each of these functions contributes a peak in the transformed 
potential, such that it features two peaks. Note that this does not depend on the particular value 
of $k_y$, chosen from (\ref{nreal}). If exactly one of the transformation functions is regular and associated with a value 
for the parameter $n$, then the transformed potential features $n+1$ peaks of finite height. This is so because the 
nonregular transformation function contributes a peak, while its regular counterpart contributes $n$ peaks.

\subsection{Higher-order SUSY transformations} 
We will now see that the effect of second-order SUSY transformations on the initial potential (\ref{v}) can be 
generalized to the higher-order case in a straightforward way. As in the previous sections we distinguish 
between regular and nonregular transformation functions. After presenting two examples we will be able to 
derive a general conclusion on the effect of an arbitrary-order SUSY transformation on the initial potential (\ref{v}). 
For the sake of brevity we will omit to show probability densities associated with the solutions of our transformed Dirac 
equation.

\begin{figure}[h]
\begin{center}
\epsfig{file=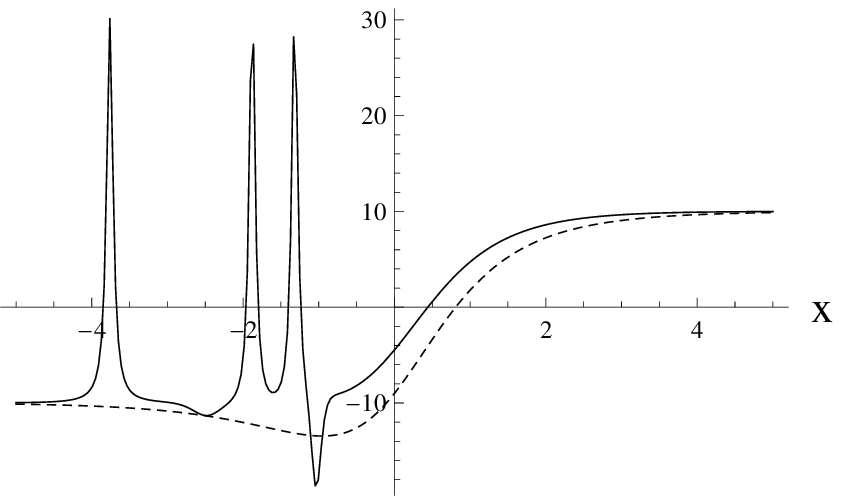,width=7.8cm}
\epsfig{file=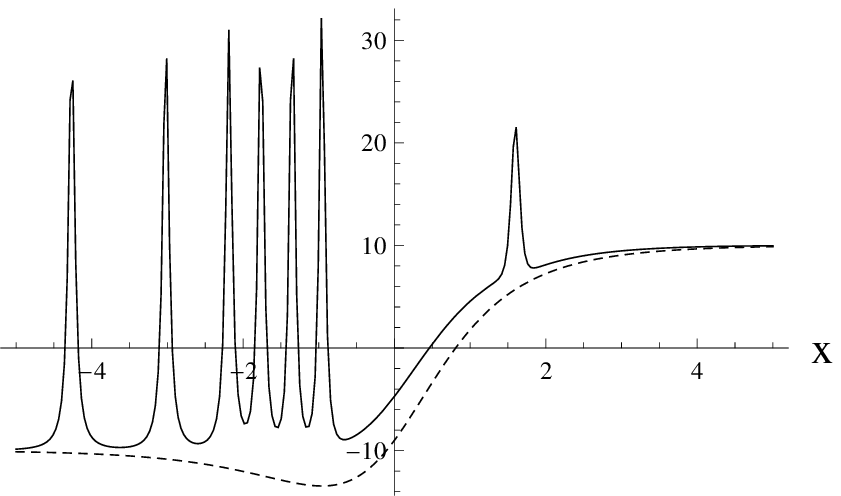,width=7.8cm}
\caption{The initial Dirac potential (\ref{v}) (dashed curve) and its 
transformed counterpart (\ref{uuse}) (solid curve), obtained from a fourth-order SUSY transformation using 
(\ref{psi1bound}) and the settings $\lambda=9$, $\mu=10$. Transformation functions are given in 
(\ref{ususy4}).}
\label{s4pot}
\end{center}
\end{figure}

\paragraph{Regular transformation functions.} Let us first present an example of a 
fourth-order SUSY transformation for parameter settings $c_1=1$, $c_2=0$, $\lambda=9$ and $\mu=10$. 
We choose four regular transformation functions from (\ref{psi1bound}) as follows
\begin{eqnarray}
u_1(x) = \Psi_{1,b}(x)_{|n=3} \qquad u_2(x) = \Psi_{1,b}(x)_{|n=6}  \qquad u_3(x) = \Psi_{1,b}(x)_{|n=7} 
 \qquad u_4(x) = \Psi_{1,b}(x)_{|n=8}. \nonumber 
\end{eqnarray}
\vspace{-.5cm}
\begin{eqnarray}
\label{ususy4}
\end{eqnarray}
We omit to show the explicit form of these functions. Their factorization energies can be obtained from 
(\ref{hole}) by insertion of the appropriate values for $n$. We obtain
\begin{eqnarray}
\epsilon_1 &=&-\left(k_y^2\right)_{|n=1}~=~-\frac{625}{4} \nonumber \\[1ex]
\epsilon_2 &=& -\left(k_y^2\right)_{|n=4}~=~-\frac{481}{4} \nonumber \\[1ex]
\epsilon_3 &=& -\left(k_y^2\right)_{|n=6}~=~-\frac{425}{4} \nonumber \\[1ex]
\epsilon_4 &=& -\left(k_y^2\right)_{|n=8}~=~-\frac{401}{4}. \label{t4}
\end{eqnarray}
We briefly verify if our reality condition (\ref{realfin}) is satisfied. To this end, we evaluate the condition for 
the present parameter settings and the transformation functions (\ref{t4}). We obtain
\begin{eqnarray}
\frac{\left|W_{u_1,u_2,u_3,u_4,\Psi_{1,b}}(x)_{|n=9/2-6i} \right|}{\left|W_{u_1,u_2,u_3,u_4}(x) \right|} 
~\approx~14146.8. \nonumber
\end{eqnarray}
Since this quantity does not depend on $x$, we are guaranteed that our transformed Dirac potential takes 
real values. We construct this potential by using relation (\ref{susy}) that reads 
\begin{eqnarray}
\Phi_1(x) &=& \frac{W_{u_1,u_2,u_3,u_4,\Psi_{1,b}}(x)}{W_{u_1,u_2,u_3,u_4}(x)}, \label{susy4n}
\end{eqnarray}
where the transformation functions are given by (\ref{ususy4}). In the next step we can calculate the 
transformed potential (\ref{uuse}) by means of 
\begin{eqnarray}
U(x) &=& \left\{\frac{W_{u_1,u_2,u_3,u_4}(x)}{W_{u_1,u_2,u_3,u_4\Psi_{1,b}}(x)} ~\frac{d}{dx} 
\left[\frac{W_{u_1,u_2,u_3,u_4,\Psi_{1,b}}(x)}{W_{u_1,u_2,u_3,u_4}(x)}\right] \right\}_{\Big| n=\frac{17}{2}-10i}. \label{ux4}
\end{eqnarray}
Note that the value for $n$ at which we evaluate the latter expression is obtained from the general 
expression $n=\lambda-i\mu-1/2$. Recall that this value for $n$ guarantees $k_y=0$. The graph of our 
transformed Dirac potential can be inspected in the left plot of figure \ref{s4pot}. We see that the fourth-order SUSY 
transformation added three finite-height peaks to the initial potential (\ref{v}). After the next example we will 
explain the behaviour of the transformed potential in a more general context.

\paragraph{Nonregular transformation functions.} In this final example we will perform a SUSY transformation 
of fourth order, where we employ three regular and one nonregular transformation function. Applying the 
parameter settings $c_1=1$, $c_2=0$, $\lambda=9$ and $\mu=10$, we define our transformation functions 
as follows
\begin{eqnarray}
u_1(x) = \Psi_1(x)_{|k_y=29/2} \qquad u_2(x) = \Psi_{1,b}(x)_{|n=3}  \qquad u_3(x) = \Psi_{1,b}(x)_{|n=4} 
 \qquad u_4(x) = \Psi_{1,b}(x)_{|n=8}. \nonumber
\end{eqnarray}
\vspace{-.5cm}
\begin{eqnarray}
\label{ususy4n}
\end{eqnarray}
We observe that the function $u_1$ is nonregular. In order to generate a real-valued transformed Dirac potential, we 
assigned a value for $k_y$ that is given by (\ref{nreal}) for $n=-2$. The functions in (\ref{ususy4n}) are associated 
with factorization energies that we get from (\ref{hole}) by substituting the values for $n$ stated in (\ref{ususy4n}). 
\begin{eqnarray}
\epsilon_1 &=&-\left(k_y^2\right)_{|n=-1}~=~-\frac{841}{4} \nonumber \\[1ex]
\epsilon_2 &=& -\left(k_y^2\right)_{|n=3}~=~-\frac{521}{4} \nonumber \\[1ex]
\epsilon_3 &=& -\left(k_y^2\right)_{|n=4}~=~-\frac{481}{4} \nonumber \\[1ex]
\epsilon_4 &=& -\left(k_y^2\right)_{|n=8}~=~-\frac{401}{4}. \label{t4n}
\end{eqnarray}
In order to assure that our transformed Dirac potential is real-valued, we evaluate our reality condition 
(\ref{realfin}) for the present parameter settings and the transformation functions (\ref{ususy4n}). 
This yields
\begin{eqnarray}
\frac{\left|W_{u_1,u_2,u_3,u_4,\Psi_{1,b}}(x)_{|n=9/2-6i} \right|}{\left|W_{u_1,u_2,u_3,u_4}(x) \right|} 
~\approx~ 18169.4. \nonumber
\end{eqnarray}
Since this quantity is constant, our reality condition is satisfied. Our transformed potential can now be calculated 
by means of the relations (\ref{susy4n}) and (\ref{ux4}). The right plot of figure \ref{s4pot} shows the graph of 
the potential. We observe that the effect of our fourth-order SUSY transformation consists in the addition of 
seven finite-height peaks to the initial potential (\ref{v}).
\paragraph{General behaviour of the transformed potential.}
As a conclusion of the preceding examples, let us now generalize the behaviour of the transformed Dirac potential 
(\ref{uuse}) to arbitrary-order SUSY transformations. Let us send ahead that subsequent statements are not obtained 
through a rigorous mathematical proof, but rather by numerical studies of higher-order SUSY partners to our 
initial potential (\ref{v}). While a few of these partners were presented in the preceding sections, numerical results 
indicate that the qualitative behavior of the SUSY-transformed potential can be generally summarized in the 
following way. Assume that our SUSY transformation 
uses $M$ regular and $N$ nonregular transformation functions. Furthermore we 
assume that the system's parameters are chosen such that our reality condition (\ref{realfin}) is satisfied, that is, 
we are guaranteed to generate a real-valued Dirac potential. Next, let $n_1,n_2,...,n_M$ be the values of $n$ that 
are associated with the $M$ regular transformation functions. For the sake of simplicity we 
assume that the list is sorted in ascending order. Then, the number of peaks that the transformed Dirac 
potential (\ref{uuse}) will feature as a result of our SUSY transformation, is calculated as follows
\begin{eqnarray}
\mbox{Number of peaks} &=& 
\left\{
\begin{array}{llll}
{\displaystyle{N-\frac{M}{2}+\sum\limits_{j=1}^{\frac{M}{2}} n_{2j}-n_{2j-1}}}
& \mbox{if $M$ is even} \\[1ex]
{\displaystyle{N+n_1-\frac{M-1}{2}+\sum\limits_{j=1}^{\frac{M-1}{2}} n_{2j+1}-n_{2j}}}
& \mbox{if $M$ is odd} \\[1ex]
\end{array}
\right\}. \label{formula}
\end{eqnarray}
Let us verify this formula by applying it to the transformation functions defined in (\ref{t4n}). We have 
$N=1$, $M=3$, $n_1=3$, $n_2=4$, $n_3=8$. Since $M$ is odd, we must use the second formula in (\ref{formula}). 
Upon substitution of the aforementioned values we find
\begin{eqnarray}
\mbox{Number of peaks} &=& 1+3-1+8-4 ~=~7. \nonumber
\end{eqnarray}
This result is correct, as can be verified by inspection of the right plot in figure \ref{s4pot}. Let us now comment on the 
interpretation of our statement (\ref{formula}) from a physical viewpoint. The main effect that our SUSY transformations 
have on the initial hyperbolic potential (\ref{v}) is the addition of finite-height peaks. The number of these peaks, their 
height, shape and location can be controlled by adjusting parameters of the SUSY transformations. The peaks 
model localized interactions that complement the initial potential. In particular, high, but very narrow peaks 
approximate delta-like perturbations.  Potentials including delta-functions have been considered in 
a variety of applications within the general context of electron transport in graphene, see for example 
\cite{barbier} \cite{hsu} and references therein.

\section{Concluding remarks}
We have devised a method for generating solutions to the massless zero-energy Dirac equation (\ref{dirac}) through application 
of the SUSY formalism. While our results feature SUSY partners of the specific hyperbolic potential (\ref{v}), the method 
works for any potential that renders the Dirac equation (\ref{dirac}) in a solvable form. Examples of such potentials are given 
in \cite{ho}, note that several of them are special cases of (\ref{v}). The general purpose of constructing our SUSY partners 
to the initial Dirac potential (\ref{v}) is twofold. First, it is of interest to identify potentials that allow for exact Dirac solutions 
because not many of such solutions have been reported in the literature so far. Particularly solutions of bound-state 
type are of interest because these can be interpreted as wavefunctions of confined electrons or holes. Second, the new 
Dirac potentials obtained from SUSY transformations can be used to approximate interactions that can be realized in 
experiments. As a further point let us recall that the equation (\ref{dirac}) we focus on in this work is the massless 
Dirac equation. A generalization of our method to the massive Dirac equation is contingent on the solvability of 
the system (\ref{syst1}), (\ref{syst2}). An additive mass parameter in this system will result in technical problems 
when trying to decouple it. We face this type of technical problem also when considering our Dirac equation (\ref{dirac}) 
at nonzero stationary energy. In contrast to the latter situation, inclusion of an effective mass function that depends on 
the spatial variables might allow adjustment to warrant decoupling of the system (\ref{syst1}), (\ref{syst2}).

\end{sloppypar}

\end{document}